\documentclass[aps,prd,twocolumn,groupedaddress,showpacs]{revtex4}
\usepackage{amsfonts,amssymb,amscd,amsmath}
\usepackage{graphicx}
\usepackage{epsfig}
\usepackage{latexsym}
\usepackage{amssymb}
\usepackage{setspace}

\begin{document}

\def\abs#1{ \left| #1 \right| }
\def\lg#1{ | #1 \rangle }
\def\rg#1{ \langle #1 | }
\def\lrg#1#2#3{ \langle #1 | #2 | #3 \rangle }
\def\lr#1#2{ \langle #1 | #2 \rangle }
\def\me#1{ \langle #1 \rangle }

\newcommand{\bra}[1]{\left\langle #1 \right\vert}
\newcommand{\ket}[1]{\left\vert #1 \right\rangle}
\newcommand{\bx}{\begin{matrix}}
\newcommand{\ex}{\end{matrix}}
\newcommand{\be}{\begin{eqnarray}}
\newcommand{\ee}{\end{eqnarray}}
\newcommand{\nn}{\nonumber \\}
\newcommand{\no}{\nonumber}
\newcommand{\de}{\delta}
\newcommand{\lt}{\left\{}
\newcommand{\rt}{\right\}}
\newcommand{\lx}{\left(}
\newcommand{\rx}{\right)}
\newcommand{\lz}{\left[}
\newcommand{\rz}{\right]}
\newcommand{\inx}{\int d^4 x}
\newcommand{\pu}{\partial_{\mu}}
\newcommand{\pv}{\partial_{\nu}}
\newcommand{\au}{A_{\mu}}
\newcommand{\av}{A_{\nu}}
\newcommand{\p}{\partial}
\newcommand{\ts}{\times}
\newcommand{\ld}{\lambda}
\newcommand{\al}{\alpha}
\newcommand{\bt}{\beta}
\newcommand{\ga}{\gamma}
\newcommand{\si}{\sigma}
\newcommand{\ep}{\varepsilon}
\newcommand{\vp}{\varphi}
\newcommand{\zt}{\mathrm}
\newcommand{\bb}{\mathbf}
\newcommand{\dg}{\dagger}
\newcommand{\og}{\omega}
\newcommand{\Ld}{\Lambda}
\newcommand{\m}{\mathcal}
\newcommand{\dm}{{(k)}}

\title{Generalized Limits for Parameter Sensitivity via Quantum Ziv-Zakai Bound}
\author{Yang Gao}
\email{gaoyangchang@gmail.com} \affiliation{Department of Physics,
Xinyang  Normal University, Xinyang, Henan 464000, China}
\author{Hwang Lee}
\affiliation{Hearne Institute for Theoretical Physics and
Department of Physics and Astronomy, \\
Louisiana State University, Baton Rouge, LA 70803, USA }
\date{\today }

\begin{abstract}
We study the generalized limit for parameter sensitivity in quantum
estimation theory considering the effects of repeated and adaptive
measurements. Based on the quantum Ziv-Zakai bound, we derive some
lower bounds for parameter sensitivity when the Hamiltonian of
system is unbounded and when the adaptive measurements are
implemented on the system. We also prove that the parameter
sensitivity is bounded by the limit of the minimum detectable
parameter. In particular, we examine several known states in quantum
phase estimation with non-interacting photons, and show that they
can not perform better than Heisenberg limit in a much simpler way
with our result.
\end{abstract}

\maketitle

\section{Introduction}
Quantum theory enables us to estimate a parameter more precisely
than classical theory \cite{squeeze}. For the quantum phase
estimation, it is known that phase sensitivity with non-interacting
photons is improved from the usual shot-noise limit (SNL), namely
$\Delta \theta \simeq 1/\sqrt {\me n}$, to Heisenberg limit (HL),
namely $\Delta \theta \simeq 1/\me n$, where $\me n$ is the mean
number of photons \cite{HL}. The underlying reason is the
superposition principle that plays an essential role. Although a
photon-number state $\lg n$ is useless for estimating phase, its
superposition with the vacuum state, $(\lg 0+\lg n)/\sqrt 2$, is
optimal for phase sensitivity with $n$ available photons, i.e.
$\Delta \theta = 1/n$. More recently, there appear some counter
examples that were used to beat the HL for phase sensitivity without
limits \cite{ssw,bhl,snoon}. However, all these proposals were based
on either crude statistical arguments or non-achievable lower bound,
such as quantum Cramer-Rao (CR) bound \cite{uncert}. More careful
calculations reveal that they approach but never beat the HL.

As for the CR bound, it sets only a lower limit for phase
sensitivity, and whether it can be achieved can only be checked by
details, that is to say, sometimes the CR bound may not be
achievable. On the other hand, by splitting the total mean number of
photons $N_T=M\me n$ into $M$ independent and identical samplings
for repeated measurements, Fisher theorem tells that the CR bound in
this case is approached asymptotically as $M\to \infty$. Hence,
whether the HL, i.e. $\Delta \theta \simeq 1/N_T$, is beaten or not
becomes more tricky. It is argued in Refs. \cite{dssw,spm} that the
optimal sensitivity with maximum likelihood estimation for $N_T$
photons will occur at $M\simeq M_\zt{knee}$, where $M_\zt{knee}$ is
the turning point after which the CR bound is asymptotically
approached. Within this scheme, Ref. \cite{dssw} examined
Shapiro-Shepard-Wong (SSW) state \cite{ssw} which was proposed to
beat the HL and found that the SSW state performs even worse than
the HL. Similar arguments can also be used for other states to check
if they beat the HL. However, this scheme has its own inconvenience
because it needs cumbersome calculations, numerically or
analytically.

It is thus very convenient to have a condition that can check if the
sensitivity with a given state can achieve HL more simply, and that
is independent of the estimation scheme. It is proposed in Ref.
\cite{ou} that by loosely relating the phase sensitivity with the
minimum detectable phase shift, such a condition can be expressed as
the fidelity between two output states --- undergoing zero and the
minimal detectable phase shift $\theta_m$ respectively --- should be
significantly different from unity as $\theta_m \simeq 1 / N_T$ and
$N_T \to \infty$. This condition applies to the single measurement
of phase shift as well as the repeated ones, where the states are
taken as the direct products of states over all measurements.
However, Ref. \cite{ou} does not present a rigorous relation
indicating that the phase sensitivity is bounded by the scaling of
the minimum detectable phase shift over $N_T$. In the present paper
we obtain such a relation in general cases, and prove some lower
bounds for the parameter sensitivity in terms of quantum Ziv-Zakai
(ZZ) bound \cite{ts,gm} when the Hamiltonian of system is unbounded
and when the adaptive measurements are implemented on the system.

In Sec. II we review the quantum ZZ bound and some recent results
for the parameter sensitivity. This bound is then applied to the
more general cases in Sec. III and to obtain our main result. In
Sec. IV we discuss some known examples of quantum phase estimation
showing that they can not perform better than HL. Finally, the
summary is given in Sec. V.

\section{Quantum Ziv-Zakai bound}

In the parameter estimation theory, the ZZ bound provide a lower
bound for the parameter sensitivity, namely the root mean-square
error other than the usual CR bound \cite{ts}. The ZZ bound connects
the parameter sensitivity to the error probability in a binary
decision problem and is often tighter than CR bound in the highly
non-Gaussian regime. Both the ZZ and CR bounds can also follow
toward each other asymptotically as the number of the repeated
measurements increases to infinity.

\begin{figure}[t!]
\begin{minipage}{.3\textwidth}
\centerline{\epsfxsize 60mm \epsffile{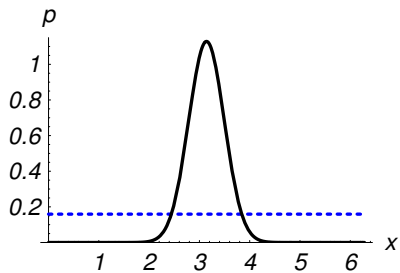}}
\end{minipage}
\caption{Comparison of the prior (dashed) and posterior (solid)
probability distributions of an unknown parameter. It indicates that
the flat prior distribution is reduced to a narrow posterior
distribution after the estimation. The prior uncertainty is thus
largely decreased into a small posterior one by obtaining more
information from the measurement results. The superiority of an
estimation scheme is quantified by the enhancement of the posterior
information over the prior one.}
\end{figure}

Let $x$ be the parameter to be estimated, $y$ be the outcome of the
measurement, and $X(y)$ be an estimator of $x$ constructed from the
outcome $y$. The parameter sensitivity of $x$ is defined as \be
\Delta Y = \lt \int dx \int dy p(y|x)p(x)[X(y)-x]^2\rt^{1/2},
\label{ppss} \ee where $p(y|x)$ is the conditional probability
distribution of obtaining a certain outcome $y$ given $x$, and
$p(x)$ is the prior probability distribution. As shown in Fig. 1,
the parameter sensitivity characterizes the uncertainty of the
posterior probability distribution after the estimation. The ZZ
bound is then given by \cite{ts,gm} \be \Delta Y &\geq & \bigg \{
\int _0^\infty d\ga \ga \int_{-\infty}^\infty dx \min
[p(x),p(x+\ga)] \nn && \times \mathrm{Pr}_{\mathrm {e}} (x,x+\ga)
\bigg \}^{1/2}, \label{zz} \ee where $\mathrm{Pr}_{\mathrm {e}}
(x,x+\ga)$ denotes the minimum error probability with equally likely
hypothesis in a binary decision problem.

In quantum parameter estimation problem, suppose the parameter $x$
be encoded in the quantum state $\rho_x$. The binary decision
problem then becomes discriminating the two possible states given by
$\rho_x$ and $\rho_{x+\ga}$ with equal prior information. For such
problem, the minimum error probability over all possible
measurements and estimations is obtained by \cite{qq}, \be
\mathrm{Pr}_{\mathrm {e}} (x,x+\ga) &=&
\frac{1}{2}[1-D(\rho_x,\rho_{x+\ga})] \nn &\geq & \frac{1}{2} \lz
1-\sqrt{1-F(\rho_x,\rho_{x+\ga})^2} \rz , \label{pre} \ee where the
distance $D$ and the fidelity $F$ for any given two states $\rho$
and $\sigma$ are defined by $D(\rho,\sigma)=\zt {tr}
\sqrt{(\rho-\sigma)^\dg (\rho-\sigma)}/2 \le 1$ and $
F(\rho,\sigma)=\zt {tr} \sqrt {\sqrt {\rho} \sigma \sqrt{\rho}}\le
1$. For pure states, $D(\psi,\chi)=\sqrt{1- F(\psi,\chi)^2}$ and $
F(\psi,\chi)=|\rg \psi \chi \rangle|$.

Assume now that $\rho_x$ is generated by the unitary evolution \be
\rho_x=e^{-i x H}\rho e^{i x H},\ee where $\rho$ is the input state
and $H$ is an effective Hamiltonian, the ground level of which is
chosen to be zero, namely $H \ge 0$. The Heisenberg limit in such
situation means that the parameter sensitivity $\Delta Y$ scales
with the average effective energy $\me H = \mathrm{tr}[H\rho]$.
Under this condition, the fidelity satisfies
$F(\rho_x,\rho_{x+\ga})=F(\rho_0,\rho_{\ga})$. For simplicity,
assume further that the prior probability distribution is a uniform
window with mean $\mu$ and width $W$ given by \be p(x)=
\frac{1}{W}\mathrm {rect}\lx \frac{x-\mu}{W} \rx, \label{prior} \ee
the standard deviation of which is thus $\Delta x=W/\sqrt{12}$.
Putting Eqs. (\ref{zz}), (\ref{pre}), and (\ref{prior}) all together
gives \be \Delta Y \geq \Delta Y_\mathrm {LB} &\equiv & \bigg \{\int
_0^{W} d\ga \ga \lx 1- \frac{\ga}{W}\rx \nn && \times \frac{1}{2}\lz
1-\sqrt{1-\mathcal{F}(\ga)^2} \rz \bigg \}^{1/2}, \label{lb} \ee
where $\m F$ is a lower bound of the fidelity $F$.

One bound for the fidelity is given by \cite{spl} $\m F(\ga) =
1-\ga\me H $ for $ 0 \le \ga \le x_0 \equiv {1/\me H}$ and $\m F =
0$ for $\ga \ge 1/\me H$. It follows from Eq. (\ref{lb}) that for $0
\le z_0 \equiv W/(2x_0) \le 1/2$, \be \Delta Y_\mathrm{LB} &=&
x_0\bigg \{ \frac{z_0^2}{3}-{15-14z_0-8z_0^2+16z_0^3 \over 48 }
\sqrt{\frac{1 - z_0}{z_0}} \nn && + \lx \pi/2 - \sin^{-1}(1 - 2z_0)
\rx \frac{5/32 - z_0/4}{z_0} \bigg \}^{1/2} \nn &\to & \Delta x
~~{\text {when}}~~ z_0 \to 0, \label{hpi} \ee and for $z_0\ge 1/2$,
\be {\Delta Y_\mathrm{LB}} &=& x_0 \bigg \{ (5/12 - \pi/8) - {1/4 -
5\pi/64 \over z_0} \bigg\}^{1/2} \nn &\to & {0.1548 \over \me H}
~~{\text {when}}~~ z_0 \to \infty. \label{lpi} \ee

Another bound for $F$ is given by \cite{spl} $\m F(\ga) = \cos
(\ga\Delta H )$ for $0 \le \ga \le x_0 \pi/2 \equiv \pi/(2\Delta H)$
and $\m F = 0$ for $\ga \ge \pi/(2\Delta H)$ with $\Delta H =
\mathrm{tr}[H^2\rho]-\me H^2$. Then Eq. (\ref{lb}) gives that for
$0\le z_0 \le \pi/4$, \be \Delta Y_\mathrm{LB} &=& x_0 \bigg \{
\frac{z_0^2}{3} + \frac{\cos(2z_0)-1}{2z_0} +
\frac{\sin(2z_0)}{2}\bigg \}^{1/2} \nn &\to & \Delta x ~~{\text
{when}}~~ z_0 \to 0, \label{11} \ee and for $z_0\ge \pi/4$, \be
\Delta Y_\mathrm{LB} &=& x_0\bigg \{ (\pi^2/16-1/2) - {1/2 -\pi/4 +
\pi^3/96 \over z_0}\bigg \}^{1/2} \nn &\to & {0.3418 \over \Delta H}
~~{\text {when}}~~ z_0 \to \infty. \label{22} \ee

\begin{figure}[t!]
\begin{minipage}{.2\textwidth}
\centerline{\epsfxsize 40mm \epsffile{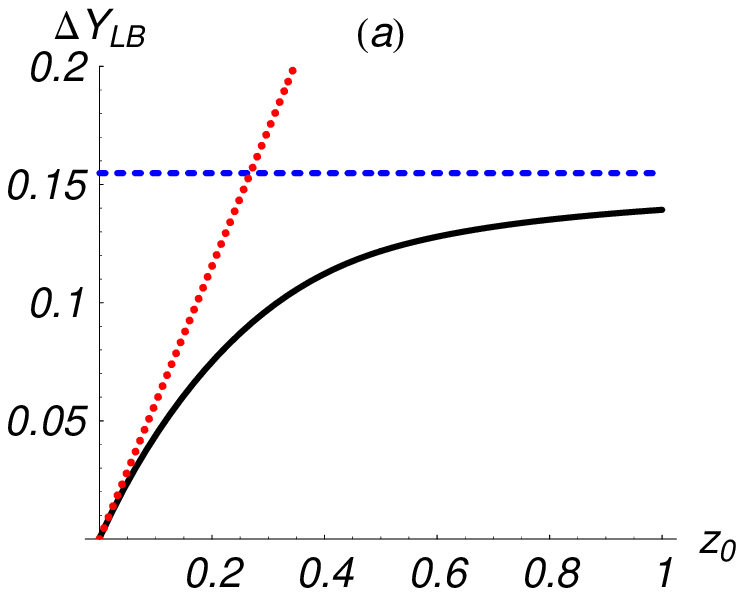}}
\end{minipage}
\begin{minipage}{.2\textwidth}
\centerline{\epsfxsize 40mm \epsffile{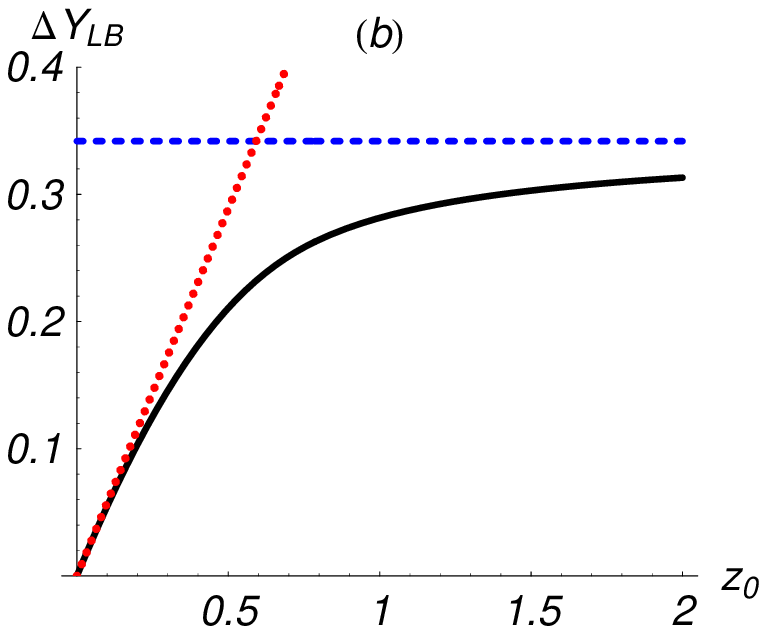}}
\end{minipage}
\caption{Comparison between the lower bounds on the parameter
sensitivity for the uniform prior distribution. (a) The solid line
is $\Delta Y_{\mathrm{LB}}$ defined in Eqs. (\ref{hpi}) and
(\ref{lpi}). The dashed line is the HL, namely $0.1548/\me H$. The
dotted line is the initial uncertainty $\Delta x = W/\sqrt{12}=x_0
z_0/\sqrt{3}$ by guessing $x$ from the prior information. (b) The
solid line is $\Delta Y_{\mathrm{LB}}$ defined in Eqs. (\ref{11})
and (\ref{22}). The dashed line is the limit $0.3418/\Delta H$. The
dotted line is the initial uncertainty $\Delta x $.  Here $\me H
=\Delta H = 1$.}
\end{figure}

As shown in Fig. 2, we can see that the parameter sensitivity
$\Delta Y$ is lower bounded by the standard deviation $\Delta x$ of
the prior distribution, i.e. \be \Delta Y \ge \Delta x, \ee when $W
\ll 1/\me H$ or $1/\Delta H$ in the high prior information (HPI)
regime, and \be \Delta Y \ge \max \lz {0.1548 \over \me H},{0.3418
\over \Delta H} \rz, \label{hhl} \ee when $W \gg 1/\me H$ or
$1/\Delta H$ in the low prior information (LPI) regime. As shown in
Fig. 2, we can see that only in the LPI regime we get the HL and the
sub-Heisenberg limit can be obtained in the HPI and intermediate
regimes. However, in the HPI regime the sub-Heisenberg strategy is
useless since one can attain the same sensitivity by just taking a
random $x$ subject to the prior distribution. On the other hand, we
can only provide a small enhancement over the initial uncertainty in
the intermediate regime by a factor of order one. Therefore, it is
not much more effective for practical estimations in the HPI and
intermediate regimes where the prior is already large enough to
allow for the sub-Heisenberg limit. Similar results can be obtained
for other prior distributions \cite{gm}.

For comparison, the quantum CR bound for the parameter sensitivity
defined by Eq. (\ref{ppss}) is \cite{qq} \be \Delta Y &\ge &
\frac{1}{\sqrt{4\Delta H^2+\Pi}},\label{fish} \ee where the quantity
$\Pi$ is the prior Fisher information, \be  \Pi & \equiv & \int dx
p(x)\lz {d \ln p(x) \over dx} \rz^2.\ee For a Gaussian prior
distribution with variance $\Delta x^2$, the Fisher information is
$\Pi=1/\Delta x^2$. From Eq. (\ref{fish}), we see that $\Delta Y \ge
\Delta x$, which is the same as the quantum ZZ bound given by Eq.
(\ref{11}) in the HPI regime, and $\Delta Y \ge 1/(2\Delta H)$,
which is more tight than the quantum ZZ bound given by Eq.
(\ref{22}) in the LPI regime.

\section{Main result}

\begin{figure}[t!]
\begin{minipage}{.3\textwidth}
\centerline{\epsfxsize 60mm \epsffile{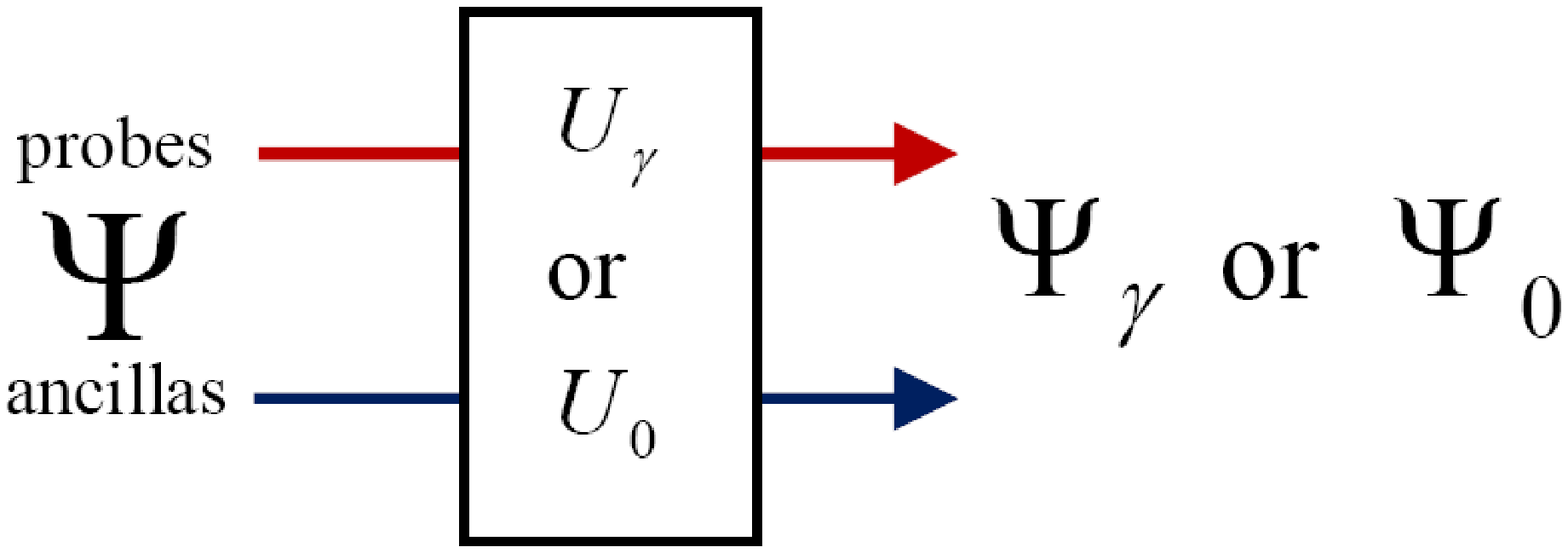}}
\end{minipage}
\caption{The schematic of single parameter estimation. The effects
of repeated and adaptive measurements are taken into considerations
by introducing ancillas and controlled unitaries. With an input
state $\Psi$, the difference between $U_0$ and $U_\ga$ is encoded in
the distinguishability of $\Psi_0$ and $\Psi_\ga$. }
\end{figure}

In the previous section we reviewed the known limits based on
quantum ZZ and CR bounds when the output state $\rho_x$ is generated
by a simple unitary $U_x(t)=e^{-i x H t}$, which does not consider
possible decoherence and measurements during the evolution interval.
In Ref. \cite{adap}, a bound for parameter sensitivity taking into
account such effect was derived via quantum CR bound only for
bounded Hamiltonian. In this section we will study the unbounded
Hamiltonian and present our main result on the generalized limit for
parameter sensitivity via quantum ZZ bound taking into account the
effect of excess decoherence, repeated and adaptive measurements
during the interval \cite{free}. Generally, the quantum dynamics of
the input state in such situation is described by completely
positive maps \cite{inf}, including sequential measurements and
feedback according to measurement outcomes.

To tackle this problem, we can first use the Kraus representation
theorem \cite{inf}, which implies that any quantum dynamics
described by completely positive maps can be reproduced by unitary
evolution of an enlarged system with appropriate ancillas, and then
use the principle of deferred measurement \cite{inf,defer,adap},
which allows us to shuffle the measurements during the evolution
time of the enlarged system to the end of the evolution time while
the measurement-based feedback is replaced by coherent controlled
unitaries prior to the overall final measurement of the enlarged
probe-ancilla system. Since our analysis below hold for all possible
measurements and estimations \cite{qq,adap} at the end of the
evolution time, we only need to consider the generalized Hamiltonian
\be H_x(t)=x H+H_0(t),\ee where the Hamiltonian $H$ contains
coupling to parameter of the probe systems, and the auxiliary
Hamiltonian $H_0$ collects all parameter-independent parts, such as
the free Hamiltonians of the probes and the controlled unitaries
induced by adaptive measurements, etc. Then, $\rho_x$ is generated
by the transformation \be \rho_x(t) = U_x(t) \rho U_x^\dg(t), \ee
where the unitary operator $U_x$ is the solution of the Schrodinger
equation \be dU_x(t)/dt=-iH_x(t)U_x(t).\ee To find out a lower bound
for the fidelity $F(\rho_x,\rho_{x+\ga})$ in this case, let $x=0$
without loss of generality, since the linear dependence of $H_x$ on
$x$.

At first, for a pure input state $\rho=\lg \Psi$, as shown in Fig.
3, the fidelity between the output states $\lg {\Psi_0}$ and $\lg
{\Psi_\ga}$ is given by \be F=\abs{\lr {\Psi_0} {\Psi_{\ga}}}= |\lrg
\Psi {U_0^\dg(t) U_\ga(t)} \Psi|.\label{fide} \ee In the interaction
picture of $H_\ga$ \cite{qm}, where $H_0$ is taken as free
Hamiltonian and $\ga H$ as interaction, we can express $U_\ga$ of
the form $ U_\ga(t)=U_0(t)\m U_\ga(t)$, where $\m U_\ga$ satisfies
the equation \be {d\m U_\ga(t)/dt}=-i \ga \m H(t) \m U_\ga(t)
\label{int} \ee with the interaction Hamiltonian $\m H = U_0^\dg H
U_0$. The solution of Eq. (\ref{int}) can be written as \be \m
U_\ga(t)= 1-i\ga \int_0^t  d s\m H(s)\m U_\ga(s). \ee So Eq.
(\ref{fide}) becomes \be F &=& |\lrg \Psi {\m U_\ga(t)} \Psi| \nn
&=& \abs { 1-i\ga \int_0^t d s \lrg \Psi {\m H(s) \m U_\ga(s)} \Psi
} \nn & \geq & 1-\ga \int_0^t d s \abs{\lrg \Psi {\m H(s) \m
U_\ga(s)} \Psi}. \label{cauchy} \ee

To proceed, we use the Cauchy inequality $ | \lr \psi \chi |^2 \leq
\lr \psi \psi \lr \chi\chi $ for $\lg \psi = \m H \lg \Psi$ and $\lg
\chi = \m U_\ga \lg \Psi$, or $\lg \psi = \sqrt{\m H} \lg \Psi$ and
$\lg \chi = \sqrt{\m H} \m U_\ga \lg \Psi$ to get \be \abs{\lrg \Psi
{\m H \m U_\ga} \Psi} \leq \min [ \sqrt{\me {\m H^2}},\sqrt{\me {\m
H } \me {\m U_\ga^\dg \m H \m U_\ga}} ] \label{cau} \ee where the
unitary of $\m U_\ga$ has been used. Let us consider two types of
$H$, depending on whether its possible energy spectra are bounded or
not. For the first type of $H$, such as in spin systems \cite{spin},
we have \be \min [ \sqrt{\me {\m H^2}},\sqrt{\me {\m H } \me {\m
U_\ga^\dg \m H \m U_\ga}} ] \le \|H\|, \label{bou} \ee where $\m H$
is transformed back to $H$ and $\|A\|=\Lambda-\lambda$ is the
semi-norm of $A$. Here $\Lambda$ ($\lambda$) is the largest
(smallest) eigenvalue of $A$. For the second type of $H$, such as in
quantum phase estimation with a coherent state, we make further
assumption that the measurements themselves do not change the energy
distributions of input state with respect to the energy spectra of
$H$, namely passive measurements, such as in the adaptive phase
estimation \cite{free} --- only the auxiliary controlled phase
shifts are introduced whereas leaving the energy distributions
untouched. Under this condition, we have \be \me {\m H^2} = \me
{H^2}, \,\ \me {\m H } =\me {\m U_\ga^\dg \m H \m U_\ga}=\me H.
\label{unbou} \ee Substituting Eqs. (\ref{bou}) and (\ref{unbou})
into Eqs. (\ref{cauchy}) and (\ref{cau}) leads to \be F &\ge & 1 -
\frac{\ga}{x_0}, \label{pu} \ee where $x_0^{-1}=\|H\|$ or $\min [
\sqrt{\langle H^2 \rangle}, \langle H \rangle ]=\langle H \rangle$
for the bounded or unbounded $H$, respectively. Here the Cauchy
inequality $\sqrt{\me {H^2}}\ge \me H$ has been used and $t=1$ is
assumed for convenience from now on.

Next, for a mixed input state $\rho=\sum_k p_k \lg {\Psi^\dm} \rg
{\Psi^\dm} $ with $\sum_k p_k =1$, we find that with $ F_k = |\lr
{\Psi_0^\dm} {\Psi_\ga^\dm}|$, \be F(\rho_0,\rho_\ga) \geq \sum_k
p_k F({\Psi^\dm_0},{\Psi^\dm_\ga}) = \sum_k p_k F_k ,\label{mini}
\ee where the convex property of fidelity \cite{inf} was used.
Putting Eq. (\ref{pu}) into Eq. (\ref{mini}), we obtain \be
F(\rho_0,\rho_\ga) \geq 1 - \frac{\ga}{x_0}. \label{gene} \ee Here
$x_0^{-1}= \|H\|$ or ${\me H}$, where the Cauchy inequality $\sum_k
p_k\sqrt{\me {H^2}_k} = \sum_k \sqrt{p_k}\sqrt{p_k\me {H^2}_k}\leq
\sqrt{\me {H^2}}$ has been used and $\me {H^2}_k =\lrg {\Psi^{(k)}}
{H^2} {\Psi^{(k)}}$. Eq. (\ref{gene}) obviously reduces to Eq.
(\ref{pu}) for pure state. Because the derivation of Eq.
({\ref{gene}) does not depend on the the assumption of $x=0$, we
thus have \be F(\rho_x,\rho_{x+\ga})\geq \m F(\ga) \label{ge}\ee for
an arbitrary $x$, where $\m F(\ga) = 1- {\ga/x_0} $ for $0\le \ga
\le x_0$ and $\m F(\ga) = 0$ for $\ga \ge x_0$. Substituting Eq.
({\ref{ge}) into Eq. ({\ref{lb}), we obtain the identical
expressions for $\Delta Y_\mathrm{LB}$ with Eqs. ({\ref{hpi}) and
({\ref{lpi}). Therefore, we get \be \Delta Y \ge \Delta x \ee in the
HPI regime, and \be \Delta Y \ge \frac{0.1548 }{\|H\|} ~{\text
{or}}~ \frac{0.1548}{\me H} \label{bd} \ee in the LPI regime for
bounded or unbounded Hamiltonian $H$ respectively. In the
intermediate regime, we can draw similar conclusions as in Sec. II.

We note that Eq. (\ref{bd}) resembles the results of Refs.
\cite{adap,fluct}. In Ref. \cite{adap} a much tighter generalized
limit based on quantum CR bound is obtained for bounded $H$, i.e.
\be \Delta Y \ge \frac{1}{\|H\|}, \ee which does not cover the cases
of quantum phase estimation with coherent and squeezed states, etc.
On the other hand, Ref. \cite{fluct} claims that \be \Delta Y \ge
\max \lz \frac{1}{M \me n},\frac{1}{\sqrt{M \me {n^2}}} \rz
\label{indep} \ee for the phase sensitivity with a linear two-mode
interferometer and $M$ repeated independent measurements. However,
it conflicts with the result of Ref. \cite{tmsv}, where the phase
sensitivity with the parity detection is found to be \be \Delta Y =
\frac{1}{\sqrt{\me n (\me n +2)}} \le \frac{1}{\me n} \ee for $M=1$,
and also fails to explain the remarkable fact in Ref. \cite{free}
where the phase sensitivity with $M$ separated photons and a proper
adaptive protocol can even achieve the HL, i.e. \be \Delta Y \approx
{4.9009 \over M} ~~{\text {for} }~~ M \gg 1. \ee For this example,
we note that $\me n = \sqrt {\me {n^2}}=1$ and Eq. (\ref{indep})
gives $\Delta Y \ge 1/\sqrt{M}$, namely the SNL. For such experiment
with $M$ separated photons, the Hamiltonian generating phase shift
can be expressed as $H= \sum_{k=1}^{M} \hat{n}_k$ with $\hat{n}_k =
a_k^\dg a_k$ being photon number operator and $a_k$ being
annihilation operator. The relevant input state can be obtained by
tracing the photon state in the reference arm $b$ over the total
state $\lg \Psi=[(\lg 0_a \lg 1_b+\lg 1_a \lg
0_b)/\sqrt{2}]^{\otimes M}$ after the first beam splitter in the
interferometer. This leads to $\rho=[(\lg 0_{aa} \rg 0+\lg 1_{aa}
\rg 1)/2]^{\otimes M}$. We thus have $\me H=M/2$ and Eq. (\ref{bd})
implies $\Delta Y \ge 0.3096/M$, namely the HL. The reason why
adaptive measurements with separated photons could achieve the HL
can be ascribed to the correlations between photons induced by
adaptive measurements, i.e. controlled unitaries.

At last, we prove that the parameter sensitivity is bounded by the
scaling limit of the minimum detectable parameter defined in Ref.
\cite{ou} versus $\me H$ for the state $\rho_x = e^{-i x H} \rho
e^{i x H}$. The minimum detectable parameter $\ga_m$ is
corresponding to the situation when the two states $\rho_{0}$ and
$\rho_{\ga_m}$ can be distinguished efficiently. Since the error
probability of discriminating the two states is
$\mathrm{Pr}_\mathrm{e}=(1-D(\rho_{0},\rho_{\ga_m}))/2$, it can be
seen that $\rho_{0}$ and $\rho_{\ga_m}$ are able to be distinguished
efficiently when $\mathrm{Pr}_\mathrm{e} \simeq 0$ and the distance
\be D(\rho_{0},\rho_{\ga_m}) \simeq 1 \ee or the fidelity \be
F(\rho_{0},\rho_{\ga_m}) \simeq 0 \label{mdp}\ee for the inequality
$D\le \sqrt{1-F^2}$. Combining Eqs. (\ref{gene}) and (\ref{mdp})
leads to \be \ga_m & \ge & x_0 = \frac{1}{\me H} \label{HL} \ee up
to some unimportant factor of order one.

Suppose the solution of Eq. (\ref{mdp}) is $\ga_m \simeq \me H
^{-\alpha}$. The lower bound provided by Eq. (\ref{HL}) puts a
constrain on the exponent $\al \le 1$. That is to say, for any
infinitesimal $\epsilon > 0 $, \be \lim_{\me H \to \infty}
F(\rho_{0},\rho_{\ga_m})|_{\ga_m \simeq \me H ^{-\alpha-\epsilon}} =
1. \label{limt} \ee In the regime of $\me H \gg 1$, we can use Eqs.
(\ref{lb}) and (\ref{limt}) to obtain \be \Delta Y_\mathrm {LB} &
\simeq  & \bigg \{\frac{1}{2}\int _0^{\min[W,\me H
^{-\alpha-\epsilon}]} d\ga \ga \lx 1- \frac{\ga}{W} \rx \bigg
\}^{1/2} \nn &=& \Delta x = \frac{W}{\sqrt{12}} \ee for $W \ll \me
H^{-\al-\epsilon}$ and \be \Delta Y_\mathrm {LB} & \simeq & \me
H^{-\al-\epsilon} \label{llbb} \ee for $W \gg \me H^{-\al-\epsilon}$
in the practical estimations. Thus, from the continuity of Eq.
(\ref{llbb}) with respect to any positive infinitesimal $\epsilon$,
we can conclude that \be \Delta Y & \ge & \mathcal{O}(\me H^{-\al})
\simeq \ga_m \label{prl}.\ee It indicates that the parameter
sensitivity $\Delta Y$ is bounded by the scaling of the minimum
detectable parameter $\ga_m$ over $\me H$ as $\me H \to \infty$.
Moreover, we see that Eq. (\ref{prl}) could be tighter than Eq.
(\ref{hhl}) since $\al \le 1$.

\section{Applications}

Some known states for quantum phase estimation that were proposed to
beat the HL have been examined in Refs. \cite{ts,gm} and found that
they can not perform better than the HL when the prior information
is appropriately considered. In the following, we use Eq.
(\ref{prl}) to re-examine these states and some other states in
Refs. \cite{bhl,snoon,tmsv} in a much simpler way. That is, we first
find out the minimum detectable phase shift $\theta_m$, and then
from Eq. (\ref{prl}) we can tell that the phase sensitivity $\Delta
\theta$ is lower bounded by the scaling of $\theta_m$ over the
average photon number in the LPI regime.

We first consider the single mode cases with $U_\theta=e^{-i\hat
n\theta}$. For the coherent state \cite{book}, $\lg {\al} = e^{\al
(a^\dg - a)}\lg 0$ with $\al$ as a real number, we can see that \be
F=\abs{\lrg \al {U_\theta} \al}= |\exp\lx {\al^2(e^{i
\theta}-1)}\rx| \simeq e^{-\me n \theta^2 /2}.\ee The minimum
detectable phase shift $\theta_m$ corresponds to the condition that
$F$ must be significantly different from unity, thus $\theta_m
\simeq 1/\sqrt{\me n}$ as $\me n =\al^2 \gg 1$, namely the SNL. If
we consider the superposition of coherent and vacuum (SCV) states,
$(\lg 0+\lg \al)/\sqrt 2$ with $\me n=\al^2/2$, we find that \be
F=|1+e^{\al^2(e^{i \theta}-1)}|/2 \simeq (1+\cos\al^2\theta)/2 \ee
and $\theta_m \simeq 1/\me n$, namely the HL. Because for coherent
state the phase factor $e^{i\al^2 \theta}$ in $\lrg \al {U_\theta}
\al$ does not contributes to $F$, while for the SCV state this term
is preserved in $F$, they give different limits for $\theta_m$. This
provides one method to construct states with higher sensitivity.

If we use coherent-squeezed (CS) state \cite{book} as input, $ \lg
{\al,r} = e^{\al (a^\dg - a)} e^{r( a^{\dg 2}-a^2)/2}\lg 0$ with
$\me n=\al^2+\sinh^2 r$ and $\Delta n=\sqrt{\al^2 e^{2r}+2\cosh^2
r\sinh^2 r}$, where the displacement $\al \gg 1$ and the squeezing
parameter $r \gg 1$, we have \be F \simeq \exp \lx
\frac{-\al^2\bt\theta^2 }{ 1+\bt^2\theta^2} \rx
/(1+\bt^2\theta^2)^{1/4},\ee where $ \bt=1/(1-\tanh r) \simeq
e^{2r}/2$. For asymptotically coherent state, $\al^2 \gg \sinh^2 r$,
\be F \simeq \exp (- e^{2r}\al^2\theta^2/2)\ee and $\theta_m \simeq
e^{-r}/\sqrt{\me n}$. On the other hand, at the optimal point $\al^2
\simeq \sinh^2 r$, \be F \simeq \exp\lx -\frac{{\me n}^2
\theta^2/2}{ 1+{\me n}^2 \theta^2} \rx / (1+{\me n}^2
\theta^2)^{1/4}\ee and $\theta_m\simeq 1/\me n$. As its two-mode
analog, one can use coherent and squeezed-vacuum state to reach the
HL in Mach-Zehnder interferometer (MZI) \cite{cohsqu}.

\begin{figure}[t!]
\begin{minipage}{.3\textwidth}
\centerline{\epsfxsize 60mm \epsffile{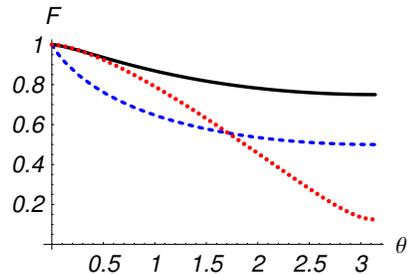}}
\end{minipage}
\caption{The plot shows the dependence of fidelity $F$ on phase
shift $\theta$ for the SSW state (dashed) $\sum_{n=0}^\Ld \lg n
/\sqrt{\zeta(2)(n+1)^2}$, dual-Fock-like state (solid)
$\sum_{n=1}^\infty {\lg n_a \lg n_b / \sqrt{\zeta(3)n^3}}$, and
noon-like state (dotted) $\sum_{n=1}^\infty \lx \lg n_a \lg 0_b+\lg
0_a \lg n_b \rx / \sqrt{2\zeta(3) n^3}$. It indicates that the
minimum detectable phase shifts with these states are of order one.}
\end{figure}

As the first proposed state to beat the HL, the SSW state is
\cite{ssw,dssw,spm} \be \lg \Psi = \frac{1}{\zeta(2)}\sum_{n=0}^\Ld
{1 \over n+1}\lg n, \ee where $\zeta(x)$ is Riemann Zeta function
and $\Ld \gg 1$. Its mean number is ${\me n} = \ln \Ld /\zeta(2)$
and variance is $\Delta n= \sqrt{\Ld/\zeta(2)}$. Here we only keep
terms up to leading order of $\Ld$. The fidelity is thus \be F\simeq
\abs{\zt {Li} _2( e^{i\theta})}/\zeta(2)\simeq 1-3\theta/\pi \ee
around $\theta=0$, where $\zt {Li} _n(x)$ is the $n$-th polynomial
logarithm. Hence the SSW state can not be used to detect a small
phase shift \cite{ou,dssw} even if ${\me n} \to \infty$, because
$F\to 1$ as $\theta \to 0$, referring to Fig. 4. However, for $M$
identical repeated measurements, we find that \be F \simeq
(1-3\theta/\pi)^M \to e^{-3/\pi} < 1 \ee as $\theta_m \simeq 1/\Ld$
and $M \simeq \Ld$. This implies the minimum detectable phase shift
with total photon number is $\theta_m \simeq (\text {logarithmic
corrections})/N_T$ associated with $N_T=M {\me n} = \Ld \ln
\Ld/\zeta(2)$

Now we examine the small peak model in Refs. \cite{bhl,spm} for
$M$-repeated measurements, $ \lg \Psi = \lg \psi^{\otimes M}$ and
$\lg \psi=(\lg 0 + \nu \lg \al)/\sqrt{1+\nu^2}$, assuming $\nu \ll
1$ and $\al\gg 1 $. Quantum CR bound gives $\Delta \theta^2 \ge
\nu^2/(M {\me n})$, where ${\me n} = \nu^2 \al^2$. In \cite{bhl} the
following parameters are chosen at will, namely $M \simeq N_T \simeq
1/\nu$ and ${\me n} \simeq 1$, then quantum CR bound leads to
$\Delta \theta^2 \ge 1/N_T^3$, and then it is claimed in Ref.
\cite{bhl} that the HL is beaten. However, as noted above, quantum
CR bound is only a lower bound and sometimes only achievable for
properly chosen parameters. If we keep ${\me n} \simeq 1$ fixed, the
fidelity is then \be F &=& \abs {\lx 1+\nu^2 e^{-\al^2(1-e^{i
\theta})}\rx / (1+\nu^2)}^M \nn &\simeq & (1-\nu^2(1-\cos \al^2
\theta))^M. \label{sp} \ee In order to make Eq. (\ref{sp}) differ
from unity significantly, we have to choose $\theta_m \simeq
1/\al^2$ and $M \simeq 1/\nu^2$, which implies that $M \simeq N_T
\simeq \al^2 $ and $1/\nu \simeq \sqrt{N_T}$, and therefore the
minimum detectable phase shift should be $\theta_m \simeq 1/N_T$.

Next, we consider the two-mode cases with field operators $a$ and
$b$, such as the MZI. For two-mode squeezed vacuum (TMSV)
\cite{tmsv}, \be \lg {\Psi}=\sqrt{1-t}\sum_n\sqrt{t^n} \lg n_a \lg
n_b \ee with $t={\me n} /(\me n +2)$, the action of the MZI is
described by unitary transformation $U_\theta=\exp \lx {\theta
(a^\dg b-b^\dg a)/2} \rx $. The fidelity is then given by \be  F &=&
(1-t)\sum_n t^n P_n(\cos \theta) \nn &=& { 1 / \sqrt{1+\me n (\me
n+2)\sin^2(\theta / 2)}} \ee in terms of Legendre polynomials $P_n$,
and thus $\theta_m \simeq 1/\sqrt{\me n(\me n+2)}$. Similarly, the
entangled coherent state was proposed to reach the HL in Ref.
\cite{ecs}, which can be expressed as $\lg \Psi = (\lg {\al}_a \lg
0_b +\lg 0_a \lg \al_b)/\sqrt{2}$ with $\me n = \al^2$ right after
the first beam splitter. The corresponding fidelity is the same as
that of the SCV state, $\theta_m \simeq 1/\me n$.

In Ref. \cite{snoon}, the following two states after the first beam
splitter in the MZI are introduced to beat the HL, i.e. noon-like
state \be \lg \Psi=\sum_{n=1}^\infty \lx \lg n_a \lg 0_b+\lg 0_a \lg
n_b \rx / \sqrt{2\zeta(3) n^3} \ee and dual-Fock-like state \be \lg
\Psi=\sum_{n=1}^\infty {\lg n_a \lg n_b / \sqrt{\zeta(3)n^3}}. \ee
It was claimed that these two states can be used to realize
unlimited phase sensitivity because they noticed $\me {n^2} \to
\infty$. However, Eq. (\ref{HL}) tells that $\theta_m \ge 1/\me n$,
which is of order one since $\me n= \zeta(2)/\zeta(3)$. Therefore,
they can not even reach the HL. If we calculate their corresponding
fidelities, \be F = \abs {1+{\zt {Li}_3(e^{i\theta})/\zeta(3) }}/2
\ee for noon-like state and \be F=\abs{\zt
{Li}_3(e^{i\theta})}/\zeta(3) \ee for dual-Fock-like state, as shown
in FIG. 4, they only differ from unity significantly at $\theta
\simeq 1$. The two states in Ref. \cite{snoon} thus can not beat the
HL.

Finally, we consider a mixed input state in the MZI \cite{tmsv},
namely \be \rho= (1-p) \lg {0,0}\rg{0,0}+ p \lg {n,n}\rg{n,n} \ee
with $\lg {n,n}=\lg n_a\lg n_b$, which has $\me n=2pn$. The distance
measure is \be D(\rho, U_\theta \rho U_\theta^\dg) &=& p\sqrt
{1-P_n^2(\cos \theta)} \nn &\simeq & p\sqrt {1-J_0^2(n \theta)}, \ee
where we have used the asymptotical expression of $P_n(\cos \theta)$
for large $n$ and $J_0(x)$ is Bessel function. In order to use this
state for efficient phase estimation, it is required that $p\simeq
1$ and then $\theta_m \simeq 1/\me n$. Otherwise, repeated
measurements are preferred. So we have verified that neither one of
the above examples can perform better than the HL.

\section{Summary}

In conclusion, we investigate the generalized limits for the
parameter sensitivity via quantum Ziv-Zakai bound, which provides a
lower bound in terms of the error probability in a quantum binary
decision problem. Such a lower bound takes into account possible
correlations induced by adaptive measurements. We also prove that
the parameter sensitivity is bounded by the scaling of the minimum
detectable parameter over the expectation value of the Hamiltonian.
At last, we examine several known states in quantum phase estimation
with non-interacting photons, and verify that neither one of them
can not perform better than the HL.

\begin{acknowledgments}
The authors wish to think J.P. Dowling for stimulating discussions.
Yang Gao would like to acknowledge support from NSFC grand No.
11147137.
\end{acknowledgments}


\end{document}